\documentclass[showpacs,aps,graphicx,twocolumn]{revtex4}
\usepackage{graphicx}

\begin{document}

\title{Efficient two-step entanglement concentration for arbitrary  W states}

\author{Yu-Bo Sheng,$^{1,2,4}$\footnote{Email address:
shengyb@njupt.edu.cn} Lan Zhou,$^3$  Sheng-Mei Zhao,$^{1,4}$ }
\address{$^1$ Institute of Signal Processing  Transmission, Nanjing
University of Posts and Telecommunications, Nanjing, 210003,  China\\
$^2$College of Telecommunications \& Information Engineering,
Nanjing University of Posts and Telecommunications,  Nanjing,
210003, China\\
 $^3$Beijing National Laboratory for Condensed Matter Physics, Institute of Physics,\\
Chinese Academy of Sciences, Beijing 100190, China\\
 $^4$Key Lab of Broadband Wireless Communication and Sensor Network
 Technology,
 Nanjing University of Posts and Telecommunications, Ministry of
 Education, Nanjing, 210003, China\\}

\begin{abstract}
We present two  two-step practical entanglement concentration
protocols (ECPs) for concentrating an arbitrary three-particle
less-entangled W state into a maximally entangled W state assisted
with single photons.   The first protocol uses the linear optics and
the second protocol adopts the cross-Kerr nonlinearity to perform
the protocol. In the first protocol,  based on the post-selection
principle, three parties say Alice, Bob and Charlie in different
distant locations can obtain the maximally entangled W state from
the arbitrary less-entangled W state with a certain success
probability. In the second protocol, it dose not require the parties
to posses the sophisticated single-photon detectors and the
concentrated photon pair can be retained after performing this
protocol successfully. Moreover,  the second protocol can be
repeated to get a higher success probability. Both protocols may be
useful in practical quantum information applications.
\end{abstract}
 \maketitle

\section{Introduction}
Entanglement is the essential role in quantum information theory
\cite{computation1,computation2,rmp}. Most of the applications of
entanglement information processing work only with the maximally
entanglement states
\cite{Ekert91,teleportation,densecoding,QSS1,QSS2,QSS3,QSDC1,QSDC2,QSDC3,QSTS1,QSTS2,QSTS3,cteleportation}.
They first produce the entanglement locally and distribute them into
different distant locations. However, during the transmission, the
particles will inevitably contact  the environment. The noisy
environment will degrade the entangled state  and make it become a
nonmaximally entangled state. Generally speaking, the nonmaximally
entangled states unusually include two  different types. The first
type is the mixed state and the second type is the pure
less-entangled state. Both of which will make the fidelity of
quantum teleportation degraded, quantum dense coding failed, and the
quantum cryptography protocol be insecure.

The method of distilling a mixed state into a maximally entangled
state is called entanglement purification, which has been studied
for several decades \cite{C.H.Bennett1,D. Deutsch,M. Murao,M.
Horodecki,Simon,Pan1,Pan2,Yong,shengpra,wangc1,wangc2,wangc3,lixhepp,dengonestep1,dengonestep2,sangouard},
since Bennett \emph{et al.} proposed an entanglement purification
protocol in 1996 \cite{C.H.Bennett1}. Another way of distilling a
pure less-entangled state into  a maximally entangled state is
called entanglement concentration, which will be detailed later
\cite{C.H.Bennett2,swapping1,swapping2,Yamamoto1,Yamamoto2,zhao1,zhao2,wangxb,bose,shengpra2,shengpra3,shengqic,shengpla,zhang,singlecopy1,he3}.
 Bennett \emph{et al.} proposed an  entanglement concentration protocol (ECP)
in 1996 \cite{C.H.Bennett2}. It is so called Schimidt projection
protocol. Bose \emph{et al.} proposed an ECP based on entanglement
swapping \cite{swapping1}. This protocol needs collective Bell-state
measurement. It was developed by Shi \emph{et al.}, subsequently
\cite{swapping2}. An ECP based on the quantum statistics was
discussed by Paunkovi\'{c} \emph{et al.}. Their protocol requires
less knowledge of the initial state than most ECPs \cite{bose}.  In
2001, Yamamoto \emph{et al.} and Zhao\emph{ et al.} proposed two
similar protocols based on linear optics \cite{Yamamoto1,zhao1}.
Their protocols were both realized experimentally
\cite{Yamamoto2,zhao2}. The ECPs based on the cross-Kerr
nonlinearity were
 proposed in 2008 \cite{shengpra2,shengqic}. However, most of the
ECPs are focused on two-particle system. They are used to
concentrate a two-particle entangled state
$\alpha|00\rangle+\beta|11\rangle$ to the Bell state
$\frac{1}{\sqrt{2}}(|00\rangle+|11\rangle)$. In a three-particle
system, there are two classes of tripartite-entangled states which
cannot be converted into each other by stochastic local operations
and classical communication \cite{dur}. They are
Greenberger-Horne-Zeilinger (GHZ) state and W state. The GHZ state
can be described as
$|GHZ\rangle=\frac{1}{\sqrt{2}}(|000\rangle+|111\rangle)$ and the W
state can be written as
$|W\rangle=\frac{1}{\sqrt{3}}(|001\rangle+|010\rangle+|100\rangle)$.
 The ECPs described above for two-particle  Bell state are usually  suitable for the
case of multipartite GHZ state
\cite{C.H.Bennett2,Yamamoto1,zhao1,shengpra2}. Thus one does not
need to discuss the ECP for GHZ state additionally. But, this kind
of ECPs cannot deal with the case of less-entangled W state. In
2003, Cao and Yang proposed an ECP for W class state with the help
of joint unitary transformation \cite{cao}. In 2007, Zhang \emph{et
al.} proposed an ECP based on the Bell-state measurement
\cite{zhanglihua}. Both  joint unitary transformation and Bell-state
measurement are not easy to realize in current condition. In 2010,
Wang \emph{et al.} proposed an ECP for W state with linear optics
\cite{wang}. Their protocol is focused on a special kind of W state,
that is $\alpha|HHV\rangle+\beta(|HVH\rangle+|VHH\rangle)$.
Recently, Yildiz  proposed an optimal distillation of three-qubit
asymmetric W states \cite{yildiz} of the form
\begin{eqnarray}
\frac{1}{\sqrt{2}}|001\rangle+\frac{1}{2}|010\rangle+\frac{1}{2}|100\rangle,\nonumber\\
\frac{1}{2}|001\rangle+\frac{1}{2}|010\rangle+\frac{1}{\sqrt{2}}|100\rangle.
\end{eqnarray}

In this paper, we present two ECPs  for concentrating an arbitrary
three-photon W state
$\alpha|VHH\rangle+\beta|HVH\rangle+\gamma|HHV\rangle$ to a standard
maximally entangled W state
$\frac{1}{\sqrt{3}}(|VHH\rangle+|HVH\rangle)+|HHV\rangle$ with only
local operation and classical communication. Here $|H\rangle$ and
$|V\rangle$ represent the horizontal and vertical polarization of
the photon, respectively. Our protocols are quite different from
others for we are focused on the arbitrary three-particle W states.
Moreover, during concentrating, we do not require two  copies of
less-entangled states, but only need one pair of less-entangled
satate and two single photons. In this way, it is more practical and
economical. The two protocols are implemented with different optical
elements. In the former, we use the polarization beam splitter (PBS)
to perform the parity check and to achieve the whole task. In the
later, we adopt the cross-Kerr nonlinearity to construct a quantum
nondemolition (QND) measurement. With the help of the QNDs, this
protocol becomes more powerful. It does not require the
sophisticated single-photon detectors and can be repeated to get a
higher success probability.

This paper is organized as follows: In Sec. II, we describe the
first protocol with linear optics. We denote it the  PBS protocol.
We show that an arbitrary less-entangled W state can be concentrated
with a certain success probability. In Sec. III, we exploit the QNDs
to substitute the PBSs and make the ECP more feasible  and efficient
in current technology. We denote it the QND protocol. In Sec. IV, we
present a discussion and summary.

\section{W state concentration with linear optics}
 In Fig. 1, we show the basic principle
of our concentration protocol. In order to explain this protocol
clearly, we divide the whole ECP into two steps. In a practical
experiment, both  steps of the operations should be performed
simultaneously. Suppose  a pair of less-entangled W state
$|\Phi\rangle_{a1b1c1}$ is sent to Alice, Bob and Charlie.  The
photon in the spatial mode $a1$ is sent to Alice. The photon in the
spatial mode $b1$ is sent  to Bob, and  $c1$ belongs to Charlie. The
photon pair is initially in the following polarization
less-entangled state
\begin{eqnarray}
|\Phi\rangle_{a1b1c1}&=&\alpha|V\rangle_{a1}|H\rangle_{b1}|H\rangle_{c1}+\beta|H\rangle_{a1}|V\rangle_{b1}|H\rangle_{c1}\nonumber\\
&+&\gamma|H\rangle_{a1}|H\rangle_{b1}|V\rangle_{c1}.\label{W state}
\end{eqnarray}
We let $\alpha$, $\beta$ and $\gamma$ be real for simple, with
$\alpha^{2}+\beta^{2}+\gamma^{2}=1$.
\begin{figure}[!h]
\begin{center}
\includegraphics[width=9cm,angle=0]{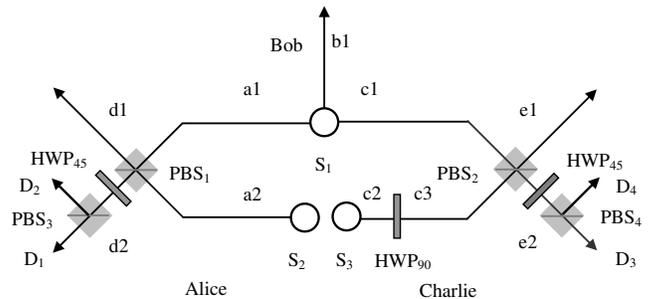}
\caption{ Schematic drawing of the first ECP with linear optics.
$S_{1}$ is the partial entanglement source and $S_{2}$ and $S_{3}$
are the single photon sources. PBSs transmit the horizontal
polarization component and reflect the vertical component.
HWP$_{90}$ and HWP$_{45}$ can rotate the polarization of the state
by $90^{\circ}$ and $45^{\circ}$, respectively.}
\end{center}
\end{figure}

In first step, a single photon in the spatial mode $a2$ emitted from
source $S_{2}$ is sent to Alice. It is described as
\begin{eqnarray}
|\Phi\rangle_{a2}&=&\frac{\alpha}{\sqrt{\alpha^{2}+\beta^{2}}}|H\rangle_{a2}+\frac{\beta}{\sqrt{\alpha^{2}+\beta^{2}}}|V\rangle_{a2}.\label{auxiliary1}
\end{eqnarray}

So the whole four-photon system can be described as
\begin{eqnarray}
|\Psi\rangle&=&|\Phi\rangle_{a1b1c1}\otimes|\Phi\rangle_{a2}=(\alpha|V\rangle_{a1}|H\rangle_{b1}|H\rangle_{c1}\nonumber\\
&+&\beta|H\rangle_{a1}|V\rangle_{b1}|H\rangle_{c1}
+\gamma|H\rangle_{a1}|H\rangle_{b1}|V\rangle_{c1})\nonumber\\
&\otimes&(\frac{\alpha}{\sqrt{\alpha^{2}+\beta^{2}}}|H\rangle_{a2}+\frac{\beta}{\sqrt{\alpha^{2}+\beta^{2}}}|V\rangle_{a2})\nonumber\\
&=&\frac{\alpha^{2}}{\sqrt{\alpha^{2}+\beta^{2}}}|V\rangle_{a1}|H\rangle_{a2}|H\rangle_{b1}|H\rangle_{c1}\nonumber\\
&+&\frac{\beta^{2}}{\sqrt{\alpha^{2}+\beta^{2}}}|H\rangle_{a1}|V\rangle_{a2}|V\rangle_{b1}|H\rangle_{c1}\nonumber\\
&+&\frac{\alpha\gamma}{\sqrt{\alpha^{2}+\beta^{2}}}|H\rangle_{a1}|H\rangle_{a2}|H\rangle_{b1}|V\rangle_{c1}\nonumber\\
&+&\frac{\beta\gamma}{\sqrt{\alpha^{2}+\beta^{2}}}|H\rangle_{a1}|V\rangle_{a2}|H\rangle_{b1}|V\rangle_{c1}\nonumber\\
&+&\frac{\alpha\beta}{\sqrt{\alpha^{2}+\beta^{2}}}|V\rangle_{a1}|V\rangle_{a2}|H\rangle_{b1}|H\rangle_{c1}\nonumber\\
&+&\frac{\alpha\beta}{\sqrt{\alpha^{2}+\beta^{2}}}|H\rangle_{a1}|H\rangle_{a2}|V\rangle_{b1}|H\rangle_{c1}.\label{combine1}
\end{eqnarray}

From Eq. (\ref{combine1}), after the two photons in spatial modes
$a1$ and $a2$ both passing through the PBS$_{1}$, the three items
$|V\rangle_{a1}|H\rangle_{a2}|H\rangle_{b1}|H\rangle_{c1}$,
$|H\rangle_{a1}|V\rangle_{a2}|V\rangle_{b1}|H\rangle_{c1}$ and
$|H\rangle_{a1}|V\rangle_{a2}|H\rangle_{b1}|V\rangle_{c1}$ will lead
to the two photons in  spatial modes $a1$ and $a2$  in the same
output mode. But the other three items
$|H\rangle_{a1}|H\rangle_{a2}|H\rangle_{b1}|V\rangle_{c1}$,
$|V\rangle_{a1}|V\rangle_{a2}|H\rangle_{b1}|H\rangle_{c1}$, and
$|H\rangle_{a1}|H\rangle_{a2}|V\rangle_{b1}|H\rangle_{c1}$ will lead
to the two output modes of PBS$_{1}$ both containing exactly one and
only one photon. Therefore, if they choose the case that the spatial
modes $d1,d2,b1$ and $c1$ all contain exactly one photon, then the
initial state  collapses to
\begin{eqnarray}
|\Psi\rangle'&=&\frac{\alpha\gamma}{\sqrt{\alpha^{2}+\beta^{2}}}|H\rangle_{d1}|H\rangle_{d2}|H\rangle_{b1}|V\rangle_{c1}\nonumber\\
&+&\frac{\alpha\beta}{\sqrt{\alpha^{2}+\beta^{2}}}|V\rangle_{d1}|V\rangle_{d2}|H\rangle_{b1}|H\rangle_{c1}\nonumber\\
&+&\frac{\alpha\beta}{\sqrt{\alpha^{2}+\beta^{2}}}|H\rangle_{d1}|H\rangle_{d2}|V\rangle_{b1}|H\rangle_{c1},\label{collpase1}
\end{eqnarray}
with a success probability of
\begin{eqnarray}
P^{1}=\frac{\alpha^{2}(\gamma^{2}+2\beta^{2})}{\alpha^{2}+\beta^{2}}.
\end{eqnarray}
The superscription "1" means the first concentration step.

Eq. (\ref{collpase1}) can be rewritten as
\begin{eqnarray}
|\Psi\rangle'&=&\frac{\gamma}{\sqrt{\gamma^{2}+2\beta^{2}}}|H\rangle_{d1}|H\rangle_{d2}|H\rangle_{b1}|V\rangle_{c1}\nonumber\\
&+&\frac{\beta}{\sqrt{\gamma^{2}+2\beta^{2}}}|V\rangle_{d1}|V\rangle_{d2}|H\rangle_{b1}|H\rangle_{c1}\nonumber\\
&+&\frac{\beta}{\sqrt{\gamma^{2}+2\beta^{2}}}|H\rangle_{d1}|H\rangle_{d2}|V\rangle_{b1}|H\rangle_{c1}.\label{collpase2}
\end{eqnarray}
Now Alice uses $\lambda/4$-wave plate HWP$_{45}$ to rotate the
photon in spatial mode $d1$. The unitary transformation of
$45^{\circ}$ rotation can be described as
\begin{eqnarray}
|H\rangle_{d2}\rightarrow\frac{1}{\sqrt{2}}(|H\rangle_{d2}+|V\rangle_{d2}),\nonumber\\
|V\rangle_{d2}\rightarrow\frac{1}{\sqrt{2}}(|H\rangle_{d2}-|V\rangle_{d2}).
\end{eqnarray}

After the rotation,  $|\Psi\rangle'$ can be written as
\begin{eqnarray}
|\Psi\rangle''&=&\frac{1}{\sqrt{2}}(\frac{\gamma}{\sqrt{\gamma^{2}+2\beta^{2}}}|H\rangle_{d1}|H\rangle_{b1}|V\rangle_{c1}\nonumber\\
&+&\frac{\beta}{\sqrt{\gamma^{2}+2\beta^{2}}}|V\rangle_{d1}|H\rangle_{b1}|H\rangle_{c1}\nonumber\\
&+&\frac{\beta}{\sqrt{\gamma^{2}+2\beta^{2}}}|H\rangle_{d1}|V\rangle_{b1}|H\rangle_{c1})|H\rangle_{d2}\nonumber\\
&+&\frac{1}{\sqrt{2}}(\frac{\gamma}{\sqrt{\gamma^{2}+2\beta^{2}}}|H\rangle_{d1}|H\rangle_{b1}|V\rangle_{c1}\nonumber\\
&-&\frac{\beta}{\sqrt{\gamma^{2}+2\beta^{2}}}|V\rangle_{d1}|H\rangle_{b1}|H\rangle_{c1}\nonumber\\
&+&\frac{\beta}{\sqrt{\gamma^{2}+2\beta^{2}}}|H\rangle_{d1}|V\rangle_{b1}|H\rangle_{c1})|V\rangle_{d2}.
\end{eqnarray}
Therefore, if the photon in spatial mode $d2$ is $|H\rangle_{d2}$,
and makes detector D$_{1}$ fire, the original state is left in the
state
\begin{eqnarray}
|\Phi_{1}\rangle_{d1b1c1}&=&\frac{\gamma}{\sqrt{\gamma^{2}+2\beta^{2}}}|H\rangle_{d1}|H\rangle_{b1}|V\rangle_{c1}\nonumber\\
&+&\frac{\beta}{\sqrt{\gamma^{2}+2\beta^{2}}}|V\rangle_{d1}|H\rangle_{b1}|H\rangle_{c1}\nonumber\\
&+&\frac{\beta}{\sqrt{\gamma^{2}+2\beta^{2}}}|H\rangle_{d1}|V\rangle_{b1}|H\rangle_{c1}.\label{concentrated1}
\end{eqnarray}

Otherwise, if  D$_{2}$ fires, the original state is left in the
state
\begin{eqnarray}
|\Phi_{2}\rangle_{d1b1c1}&=&\frac{\gamma}{\sqrt{\gamma^{2}+2\beta^{2}}}|H\rangle_{d1}|H\rangle_{b1}|V\rangle_{c1}\nonumber\\
&-&\frac{\beta}{\sqrt{\gamma^{2}+2\beta^{2}}}|V\rangle_{d1}|H\rangle_{b1}|H\rangle_{c1}\nonumber\\
&+&\frac{\beta}{\sqrt{\gamma^{2}+2\beta^{2}}}|H\rangle_{d1}|V\rangle_{b1}|H\rangle_{c1}.\label{concentrated2}
\end{eqnarray}

In order to get $|\Phi_{1}\rangle_{d1b1c1}$, one of   the parties,
says Alice, Bob or Charlie should perform a local operation of phase
rotation on her or his photon.  It is the first step of the first
ECP.

The second step  is analogy with the first one.  It is performed by
Charlie, shown in Fig. 1. After they get the state
$|\Phi_{1}\rangle_{d1b1c1}$, another single photon state
$|\Phi\rangle_{c2}$ emitted from source $S_{3}$ is sent to Charlie.
$|\Phi\rangle_{c2}$ can be written as
\begin{eqnarray}
|\Phi\rangle_{c2}&=&\frac{\beta}{\sqrt{\gamma^{2}+\beta^{2}}}|H\rangle_{c2}+\frac{\gamma}{\sqrt{\gamma^{2}+\beta^{2}}}|V\rangle_{c2}.\label{auxiliary2}
\end{eqnarray}
 Charlie first rotates the
photon by $90^{\circ}$  in the spatial mode $c2$ with HWP$_{90}$.
The $|\Phi\rangle_{c2}$ can be written as
\begin{eqnarray}
|\Phi\rangle_{c3}&=&\frac{\beta}{\sqrt{\gamma^{2}+\beta^{2}}}|V\rangle_{c3}+\frac{\gamma}{\sqrt{\gamma^{2}+\beta^{2}}}|H\rangle_{c3}.\label{auxiliary3}
\end{eqnarray}

 The combination of four-photon state can be written as
\begin{eqnarray}
|\Phi\rangle_{c3}&\otimes&|\Phi_{2}\rangle_{d1b1c1}\nonumber\\
&=&\frac{\beta\gamma}{\sqrt{\gamma^{2}+2\beta^{2}}\sqrt{\gamma^{2}+\beta^{2}}}|H\rangle_{d1}|H\rangle_{b1}|V\rangle_{c1}|V\rangle_{c3}\nonumber\\
&+&\frac{\gamma^{2}}{\sqrt{\gamma^{2}+2\beta^{2}}\sqrt{\gamma^{2}+\beta^{2}}}|H\rangle_{d1}|H\rangle_{b1}|V\rangle_{c1}|H\rangle_{c3}\nonumber\\
&+&\frac{\beta^{2}}{\sqrt{\gamma^{2}+2\beta^{2}}\sqrt{\gamma^{2}+\beta^{2}}}|V\rangle_{d1}|H\rangle_{b1}|H\rangle_{c1}|V\rangle_{c3}\nonumber\\
&+&\frac{\beta\gamma}{\sqrt{\gamma^{2}+2\beta^{2}}\sqrt{\gamma^{2}+\beta^{2}}}|V\rangle_{d1}|H\rangle_{b1}|H\rangle_{c1}|H\rangle_{c3}\nonumber\\
&+&\frac{\beta^{2}}{\sqrt{\gamma^{2}+2\beta^{2}}\sqrt{\gamma^{2}+\beta^{2}}}|H\rangle_{d1}|V\rangle_{b1}|H\rangle_{c1}|V\rangle_{c3}\nonumber\\
&+&\frac{\beta\gamma}{\sqrt{\gamma^{2}+2\beta^{2}}\sqrt{\gamma^{2}+\beta^{2}}}|H\rangle_{d1}|V\rangle_{b1}|H\rangle_{c1}|H\rangle_{c3}.\nonumber\\\label{collpase3}
\end{eqnarray}

It is easy to find that the three items
$|H\rangle_{d1}|H\rangle_{b1}|V\rangle_{c1}|H\rangle_{c3}$,
$|V\rangle_{d1}|H\rangle_{b1}|H\rangle_{c1}|V\rangle_{c3}$ and
$|H\rangle_{d1}|V\rangle_{b1}|H\rangle_{c1}|V\rangle_{c3}$ will lead
to the two photons in Charlie's location  in the same output mode
after passing through the PBS$_{2}$. But the other three items
$|H\rangle_{d1}|H\rangle_{b1}|V\rangle_{c1}|V\rangle_{c3}$,
$|V\rangle_{d1}|H\rangle_{b1}|H\rangle_{c1}|H\rangle_{c3}$ and
$|H\rangle_{d1}|V\rangle_{b1}|H\rangle_{c1}|H\rangle_{c3}$ will lead
to the two photons in different output modes. Therefore, similar to
the first step, Charlie chooses the case that two output modes of
PBS$_{2}$ both contain one photon. So  Eq. (\ref{collpase3}) becomes
\begin{eqnarray}
|\Psi'''\rangle&=&\frac{1}{\sqrt{3}}(|H\rangle_{d1}|H\rangle_{b1}|V\rangle_{e1}|V\rangle_{e2}\nonumber\\
&+&|V\rangle_{d1}|H\rangle_{b1}|H\rangle_{e1}|H\rangle_{e2}+|H\rangle_{d1}|V\rangle_{b1}|H\rangle_{e1}|H\rangle_{e2}),\nonumber\\
\end{eqnarray}
with a success probability of
\begin{eqnarray}
P^{2}=\frac{3\beta^{2}\gamma^{2}}{(\gamma^{2}+\beta^{2})(\gamma^{2}+2\beta^{2})}.
\end{eqnarray}
The superscription "2" means the second concentration step.

Finally, Charlie rotates his photon in the mode $e2$ by $45^{\circ}$
with HWP$_{45}$ and makes
\begin{eqnarray}
|H\rangle_{e2}\rightarrow\frac{1}{\sqrt{2}}(|H\rangle_{e2}+|V\rangle_{e2}),\nonumber\\
|V\rangle_{e2}\rightarrow\frac{1}{\sqrt{2}}(|H\rangle_{e2}-|V\rangle_{e2}).
\end{eqnarray}
If D$_{3}$ fires, they will get
\begin{eqnarray}
|\Phi_{1}\rangle_{d1b1e1}&=&\frac{1}{\sqrt{3}}(|H\rangle_{d1}|H\rangle_{b1}|V\rangle_{e1}\nonumber\\
&+&|V\rangle_{d1}|H\rangle_{b1}|H\rangle_{e1}+|H\rangle_{d1}|V\rangle_{b1}|H\rangle_{e1}).\nonumber\\\label{concentrated3}
\end{eqnarray}
If D$_{4}$ fires, they will get
\begin{eqnarray}
|\Phi_{2}\rangle_{d1b1e1}&=&\frac{1}{\sqrt{3}}(-|H\rangle_{d1}|H\rangle_{b1}|V\rangle_{e1}\nonumber\\
&+&|V\rangle_{d1}|H\rangle_{b1}|H\rangle_{e1}+|H\rangle_{d1}|V\rangle_{b1}|H\rangle_{e1}).\nonumber\\\label{concentrated4}
\end{eqnarray}

Both  Eqs. (\ref{concentrated3}) and (\ref{concentrated4}) are the
maximally entangled W states. In order to get
$|\Phi_{1}\rangle_{d1b1e1}$, one of  three e parties, says Alice,
Bob or Charlie should perform a local operation of phase rotation on
her or his photon.

Thus far, we have fully explained our first ECP. The total success
probability $P_{s}$ for obtaining a maximally entangled W state is
\begin{eqnarray}
P_{s}=P^{1}P^{2}&=&\frac{\alpha^{2}(\gamma^{2}+2\beta^{2})}{\alpha^{2}+\beta^{2}}\frac{3\beta^{2}\gamma^{2}}{(\gamma^{2}+\beta^{2})(\gamma^{2}+2\beta^{2})}\nonumber\\
&=&\frac{3\alpha^{2}\beta^{2}\gamma^{2}}{(\alpha^{2}+\beta^{2})(\gamma^{2}+\beta^{2})}.
\end{eqnarray}
Actually, during the whole procedure, Alice and Charlie use
essentially the same principle to perform the concentration
protocol. They both pick up the even parity states
$|H\rangle|H\rangle$ and $|V\rangle|V\rangle$ after the photons
passing through the PBSs and discard the odd parity states
$|H\rangle|V\rangle$ and $|V\rangle|H\rangle$ because the two
photons are in the same spatial modes. In fact, the discarded items
can also be reused to obtain the maximally entangled W states. So
this kind of protocol is a not optimal one.

In the above description, we explain the total ECP by  dividing it
into two steps. The first protocol is essentially to obtain the
state in Eq. (\ref{concentrated1}) from Eq. (\ref{W state}), and the
second step is to obtain the genuine maximally entangled W state
from Eq.(\ref{concentrated3}). Actually, we should point out that in
a practical experiment, we cannot perform this protocol like that.
The main reason is that this kind of ECP is based on the
post-selection principle. That is to say, they should resort the
sophisticated single-photon detectors to check the photon number in
the output modes of PBSs. For instance, in the first step, the
successful case will make both of spatial modes $d1$ and $d2$
contain one photon. However, once the photons are successfully
detected, the whole photon-state is destroyed. It is impossible to
perform the further step. Thus, the feasibly way is to perform the
two steps simultaneously, and choose the cases that the spatial
modes $d1,d2,e1,e2$ and $b1$ contain exactly one and only one photon
with the success probability of $P_{s}$.

\section{W state concentration with cross-Kerr nonlinearity}

In Sec. II, we have fully explained our first ECP with linear
optics. The whole protocol should resorts sophisticated
single-photon detectors to check the photon number. Moreover, the
whole protocol is based on the post-selection principle and requires
Alice and Charlie to perform the two steps simultaneously. These
disadvantages may limit its practical application in current quantum
information processing.

In this section, we adopt the cross-Kerr nonlinearity to implement a
QND, which can play the roles of both parity check and single-photon
detector.  Before we start this protocol, let us briefly explain the
basic principle of the cross-Kerr nonlinearity. The cross-kerr
nonlinearity has been widely studied in constructing of CNOT gates
\cite{QND1}, performing
 entanglement
purification protocol \cite{shengpra},  ECPs
\cite{shengpra2,shengqic}, complete Bell-state analysis \cite{QND2},
and other quantum communication and computation processing
\cite{he1,he2,lin1,lin2,qubit1,qubit2,qubit3,shengbellstateanalysis,zhangshou}.
The Hamiltonian of a cross-Kerr nonlinear medium can be described as
$H=\hbar\chi \hat{n_{a}}\hat{n_{b}}$, where the $\hbar\chi$ is the
coupling strength of the nonlinearity. It is decided by the
cross-Kerr material. The $\hat{n_{a}}(\hat{n_{b}})$ is the number
operator for mode $a(b)$ \cite{QND1}. In Fig. 2,  two polarized
photons are initially prepared in the states
$|\varphi\rangle_{a_1}=c_{0}|H \rangle_{a_1}+c_{1}|V\rangle_{b_1}$
and
$|\varphi\rangle_{a_2}=d_{0}|H\rangle_{a_2}+d_{1}|V\rangle_{a_2}$.
They combine with a coherent beam $|\alpha\rangle_{p}$ and interact
with the cross-Kerr nonlinearities.  So the state of the composite
quantum system from the original one $\vert \Psi\rangle_{O}=\vert
\varphi\rangle_{a_1}\otimes \vert \varphi\rangle_{a_2}\otimes \vert
\alpha\rangle_{p}$ evolves to
\begin{eqnarray}
|\Psi\rangle_{T} &=& [c_{0}d_{0}|HH\rangle +
c_{1}d_{1}|VV\rangle]|\alpha e^{i\theta}\rangle_{p} \nonumber\\
&+& c_{0}d_{1}|HV\rangle|\alpha
e^{i2\theta}\rangle_{p}+c_{1}d_{0}|VH\rangle|\alpha\rangle_{p}.\label{crosskerr}
\end{eqnarray}
From Eq. (\ref{crosskerr}), the items
 $|HH\rangle$ and $|VV\rangle$ make
the coherent beam $\vert\alpha\rangle_p$ pick up a $\theta$ phase
shift. The item $|HV\rangle$ picks up a $2\theta$ phase shift, and
the item $|VH\rangle$ picks up no phase shift. With a general
homodyne-heterodyne measurement ($X$ homodyne measurement), one can
distinguish $|HH\rangle$ and $|VV\rangle$ from $|HV\rangle$ and
$|VH\rangle$ according to their different phase shift \cite{QND1}.
It plays essentially  the same role of parity check.

Now we reconsider the first step of ECP  in Sec. II. In Fig. 3, the
QND1 and QND2 are described in Fig. 2. The $|\Phi\rangle_{a1b1c1}$
and $|\Phi\rangle_{a2}$ coupled with the coherent state  evolves as
\begin{figure}[!h]
\begin{center}
\includegraphics[width=6cm,angle=0]{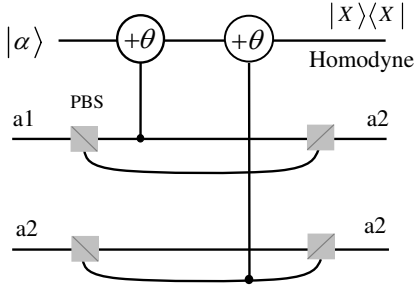}
\caption{The principle of nondestructive quantum nondemotion
detector (QND) \cite{shengpra2}. It can be used to perform the
parity check, that is to distinguish $|HH\rangle$ and $|VV\rangle$
from $|HV\rangle$ and $|VH\rangle$ according to the different phase
shift of the coherent state $|\alpha\rangle$.}
\end{center}
\end{figure}

\begin{eqnarray}
|\Psi\rangle|\alpha\rangle&=&|\Phi\rangle_{a1b1c1}\otimes|\Phi\rangle_{a2}|\alpha\rangle=(\alpha|V\rangle_{a1}|H\rangle_{b1}|H\rangle_{c1}\nonumber\\
&+&\beta|H\rangle_{a1}|V\rangle_{b1}|H\rangle_{c1}
+\gamma|H\rangle_{a1}|H\rangle_{b1}|V\rangle_{c1})\nonumber\\
&\otimes&(\frac{\alpha}{\sqrt{\alpha^{2}+\beta^{2}}}|H\rangle_{a2}+\frac{\beta}{\sqrt{\alpha^{2}+\beta^{2}}}|V\rangle_{a2})|\alpha\rangle\nonumber\\
&\rightarrow&\frac{\alpha^{2}}{\sqrt{\alpha^{2}+\beta^{2}}}|V\rangle_{a1}|H\rangle_{a2}|H\rangle_{b1}|H\rangle_{c1}|\alpha\rangle\nonumber\\
&+&\frac{\beta^{2}}{\sqrt{\alpha^{2}+\beta^{2}}}|H\rangle_{a1}|V\rangle_{a2}|V\rangle_{b1}|H\rangle_{c1}|\alpha e^{i2\theta}\rangle\nonumber\\
&+&\frac{\alpha\gamma}{\sqrt{\alpha^{2}+\beta^{2}}}|H\rangle_{a1}|H\rangle_{a2}|H\rangle_{b1}|V\rangle_{c1}|\alpha e^{i\theta}\rangle\nonumber\\
&+&\frac{\beta\gamma}{\sqrt{\alpha^{2}+\beta^{2}}}|H\rangle_{a1}|V\rangle_{a2}|H\rangle_{b1}|V\rangle_{c1}|\alpha e^{i2\theta}\rangle\nonumber\\
&+&\frac{\alpha\beta}{\sqrt{\alpha^{2}+\beta^{2}}}|V\rangle_{a1}|V\rangle_{a2}|H\rangle_{b1}|H\rangle_{c1}|\alpha e^{i\theta}\rangle\nonumber\\
&+&\frac{\alpha\beta}{\sqrt{\alpha^{2}+\beta^{2}}}|H\rangle_{a1}|H\rangle_{a2}|V\rangle_{b1}|H\rangle_{c1}|\alpha
e^{i\theta}\rangle.\label{combine3}
\end{eqnarray}
From Eq. (\ref{combine3}), after the two photons in spatial modes
$a1$ and $a2$ passing through the QND1, if Alice chooses  the
$\theta$ phase shift, the remaining state is the same as Eq.
(\ref{collpase1}), with the same probability  $P^{1}$. Then
following the same step described in Sec. II, if D$_{1}$ fires, they
will get $|\Phi_{1}\rangle_{d1b1c1}$, and if D$_{2}$ fires, they
will get $|\Phi_{2}\rangle_{d1b1c1}$.

The second step is analogy with the first one. They choose another
single photon $|\Phi\rangle_{c2}$ and then rotate it by $90^{\circ}$
with HWP$_{90}$.  The $|\Phi_{2}\rangle_{d1b1c1}$ and
$|\Phi\rangle_{c3}$ combined with the coherent state evolves as
\begin{eqnarray}
&&|\Phi\rangle_{c3}\otimes|\Phi_{2}\rangle_{d1b1c1}|\alpha\rangle\nonumber\\
&\rightarrow&\frac{\beta\gamma}{\sqrt{\gamma^{2}+2\beta^{2}}\sqrt{\gamma^{2}+\beta^{2}}}|H\rangle_{d1}|H\rangle_{b1}|V\rangle_{c1}|V\rangle_{e2}|\alpha e^{i\theta}\rangle\nonumber\\
&+&\frac{\gamma^{2}}{\sqrt{\gamma^{2}+2\beta^{2}}\sqrt{\gamma^{2}+\beta^{2}}}|H\rangle_{d1}|H\rangle_{b1}|V\rangle_{c1}|H\rangle_{e2}|\alpha \rangle\nonumber\\
&+&\frac{\beta^{2}}{\sqrt{\gamma^{2}+2\beta^{2}}\sqrt{\gamma^{2}+\beta^{2}}}|V\rangle_{d1}|H\rangle_{b1}|H\rangle_{c1}|V\rangle_{e2}|\alpha e^{i\theta}\rangle\nonumber\\
&+&\frac{\beta\gamma}{\sqrt{\gamma^{2}+2\beta^{2}}\sqrt{\gamma^{2}+\beta^{2}}}|V\rangle_{d1}|H\rangle_{b1}|H\rangle_{c1}|H\rangle_{e2}|\alpha e^{i2\theta}\rangle\nonumber\\
&+&\frac{\beta^{2}}{\sqrt{\gamma^{2}+2\beta^{2}}\sqrt{\gamma^{2}+\beta^{2}}}|H\rangle_{d1}|V\rangle_{b1}|H\rangle_{c1}|V\rangle_{e2}|\alpha e^{i2\theta}\rangle\nonumber\\
&+&\frac{\beta\gamma}{\sqrt{\gamma^{2}+2\beta^{2}}\sqrt{\gamma^{2}+\beta^{2}}}|H\rangle_{d1}|V\rangle_{b1}|H\rangle_{c1}|H\rangle_{e2}|\alpha
e^{i\theta}\rangle.\nonumber\\\label{combine4}
\end{eqnarray}
\begin{figure}[!h]
\begin{center}
\includegraphics[width=9cm,angle=0]{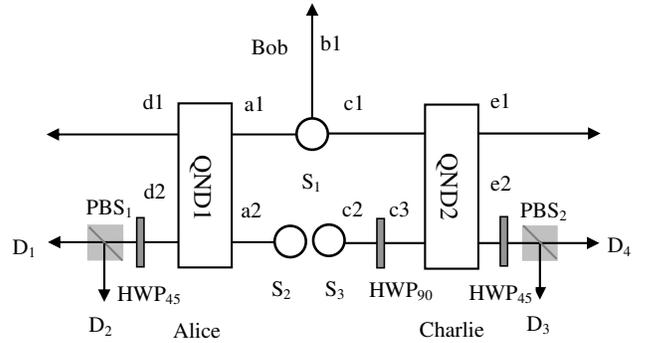}
\caption{Schematic drawing of our second ECP with the  cross-Kerr
nonlinearity. Compared with Fig. 1, we use two QNDs described in
Fig. 2 to substitute the two PBSs. It can reach a higher success
probability than the first ECP.}
\end{center}
\end{figure}

From Eq. (\ref{combine4}), if Charlie picks up the $\theta$ phase
shift, the remaining state is essentially the four-photon maximally
entangled W state. Thus, following the same way, Charlie measures
his photon in mode $e2$ after rotating it  by $45^{\circ}$ with
HWP$_{45}$. Finally, if  D$_{3}$ fires, they will get the same state
Eq. (\ref{concentrated3}). Otherwise, if the detector D$_{4}$ fires,
they will obtain the same state in Eq. (\ref{concentrated4}), with
the same probability  $P^{2}$.

Compared with the first protocol ,  the function of the QND is also
the parity check.  Certainly, with the help of the QND, we do not
need to measure the photon directly, and the concentrated photon
pairs can be remained.  During the whole process, Alice and Charlie
both pick up the cases that the phase shift is $\theta$ and discard
the other results. Interestingly, If a suitable cross-Kerr
nonlinearity can be provided, and the interaction time $t$ can be
well controlled, which leads to $\theta=\pi$. In this way, phase
shift $2\theta=2\pi$ and 0 will not be distinguished. Therefore, the
discarded items in each step by Alice and Charlie are also the
nonmaximally entangled W state and can be reconcentrated in the next
round. For instance, in Eq. (\ref{combine3}), if the phase shift is
not $\theta$ $(\pi)$, but $2\pi$ (0), that the whole state collapses
to
\begin{eqnarray}
|\Psi_{1}\rangle'_{a1a2b1c1}&=&\frac{\alpha^{2}}{\sqrt{\alpha^{2}+\beta^{2}}}|V\rangle_{a1}|H\rangle_{a2}|H\rangle_{b1}|H\rangle_{c1}\nonumber\\
&+&\frac{\beta^{2}}{\sqrt{\alpha^{2}+\beta^{2}}}|H\rangle_{a1}|V\rangle_{a2}|V\rangle_{b1}|H\rangle_{c1}\nonumber\\
&+&\frac{\beta\gamma}{\sqrt{\alpha^{2}+\beta^{2}}}|H\rangle_{a1}|V\rangle_{a2}|H\rangle_{b1}|V\rangle_{c1}.\label{less0}\nonumber\\
\end{eqnarray}
By measuring the photon in the mode $a2$ after rotating it by
$45^{\circ}$, they will get another lesser-entangled state of the
form
\begin{eqnarray}
|\Psi^{\pm}_{1}\rangle'_{a1b1c1}&=&\alpha'|V\rangle_{a1}|H\rangle_{b1}|H\rangle_{c1}
\pm\beta'|H\rangle_{a1}|V\rangle_{b1}|H\rangle_{c1}\nonumber\\
&\pm&\gamma'|H\rangle_{a1}|H\rangle_{b1}|V\rangle_{c1},\label{less1}
\end{eqnarray}
with
\begin{eqnarray}
\alpha'=\frac{\alpha^{4}}{\sqrt{\alpha^{4}+\beta^{4}+\beta^{2}\gamma^{2}}},\nonumber\\
\beta'=\frac{\beta^{4}}{\sqrt{\alpha^{4}+\beta^{4}+\beta^{2}\gamma^{2}}},\nonumber\\
\gamma'=\frac{\beta^{2}\gamma^{2}}{\sqrt{\alpha^{4}+\beta^{4}+\beta^{2}\gamma^{2}}}.
\end{eqnarray}
'+' or '-' depends on the measurement results. If D$_{1}$ fires, it
is '+', otherwise, it is '-'.

In the second step, if the phase shift in Charlie's location is not
$\theta$ yet, then the Eq. (\ref{combine4}) becomes
\begin{eqnarray}
&&|\Psi_{2}\rangle'_{d1b1c1e2}\nonumber\\
&=&\frac{\gamma^{2}}{\sqrt{\gamma^{2}+2\beta^{2}}\sqrt{\gamma^{2}+\beta^{2}}}|H\rangle_{d1}|H\rangle_{b1}|V\rangle_{c1}|H\rangle_{e2}\nonumber\\
&+&\frac{\beta^{2}}{\sqrt{\gamma^{2}+2\beta^{2}}\sqrt{\gamma^{2}+\beta^{2}}}|V\rangle_{d1}|H\rangle_{b1}|H\rangle_{c1}|V\rangle_{e2}\nonumber\\
&+&\frac{\beta^{2}}{\sqrt{\gamma^{2}+2\beta^{2}}\sqrt{\gamma^{2}+\beta^{2}}}|H\rangle_{d1}|V\rangle_{b1}|H\rangle_{c1}|V\rangle_{e2}.\label{less20}
\end{eqnarray}
By measuring the photon in mode $e2$ after rotating it by
$45^{\circ}$, it becomes
\begin{eqnarray}
|\Psi^{\pm}_{2}\rangle'_{d1b1c1}&=&\gamma''|H\rangle_{d1}|H\rangle_{b1}|V\rangle_{c1}\pm\beta''|V\rangle_{d1}|H\rangle_{b1}|H\rangle_{c1}\nonumber\\
&\pm&\beta''|H\rangle_{d1}|V\rangle_{b1}|H\rangle_{c1},\label{less2}
\end{eqnarray}
with
\begin{eqnarray}
\gamma''=\frac{\gamma^{2}}{\sqrt{\gamma^{4}+2\beta^{4}}},\nonumber\\
\beta''=\frac{\beta^{2}}{\sqrt{\gamma^{4}+2\beta^{4}}}.
\end{eqnarray}

'+' or '-' also depends on the measurement result. If D$_{3}$ fires,
it is '+', otherwise, it is '-'.

Compared with Eqs. (\ref{W state}) and (\ref{concentrated1}), it is
obvious to see that Eqs. (\ref{less1}) and (\ref{less2}) have the
same form with Eqs.(\ref{W state}) and (\ref{concentrated1}). That
is to say, the states of  Eqs. (\ref{less1}) and (\ref{less2}) can
be reconcentrated to get a maximally entangled W state in the next
round. We take $|\Psi^{+}_{2}\rangle'_{d1b1c1}$ as an example. In
detail, Charlie chooses another single photon of the form
\begin{eqnarray}
|\Phi'\rangle_{c2}=\frac{\beta^{2}}{\sqrt{\gamma^{4}+\beta^{4}}}|H\rangle_{c2}+\frac{\gamma^{2}}{\sqrt{\gamma^{4}+\beta^{4}}}|V\rangle_{c2}.
\end{eqnarray}
After rotating this photon by $90^{\circ}$, it becomes
\begin{eqnarray}
|\Phi'\rangle_{c3}=\frac{\beta^{2}}{\sqrt{\gamma^{4}+\beta^{4}}}|V\rangle_{c3}+\frac{\gamma^{2}}{\sqrt{\gamma^{4}+\beta^{4}}}|H\rangle_{c3}.
\end{eqnarray}
Therefore, states $|\Psi^{+}_{2}\rangle'_{d1b1c1}$ and
$|\Phi'\rangle_{c3}$ combined with the coherent state
$|\alpha\rangle$ evolves as
\begin{eqnarray}
&&|\Phi'\rangle_{c3}\otimes|\Psi^{+}_{2}\rangle'_{d1b1c1}|\alpha\rangle\nonumber\\
&\rightarrow&\frac{\beta^{2}\gamma^{2}}{\sqrt{\gamma^{4}+2\beta^{4}}\sqrt{\gamma^{4}+\beta^{4}}}|H\rangle_{d1}|H\rangle_{b1}|V\rangle_{c1}|V\rangle_{e2}|\alpha e^{i\theta}\rangle\nonumber\\
&+&\frac{\gamma^{4}}{\sqrt{\gamma^{4}+2\beta^{4}}\sqrt{\gamma^{4}+\beta^{4}}}|H\rangle_{d1}|H\rangle_{b1}|V\rangle_{c1}|H\rangle_{e2}|\alpha \rangle\nonumber\\
&+&\frac{\beta^{4}}{\sqrt{\gamma^{4}+2\beta^{4}}\sqrt{\gamma^{4}+\beta^{4}}}|V\rangle_{d1}|H\rangle_{b1}|H\rangle_{c1}|V\rangle_{e2}|\alpha e^{i\theta}\rangle\nonumber\\
&+&\frac{\beta^{2}\gamma^{2}}{\sqrt{\gamma^{4}+2\beta^{4}}\sqrt{\gamma^{4}+\beta^{4}}}|V\rangle_{d1}|H\rangle_{b1}|H\rangle_{c1}|H\rangle_{e2}|\alpha e^{i2\theta}\rangle\nonumber\\
&+&\frac{\beta^{4}}{\sqrt{\gamma^{4}+2\beta^{4}}\sqrt{\gamma^{4}+\beta^{4}}}|H\rangle_{d1}|V\rangle_{b1}|H\rangle_{c1}|V\rangle_{e2}|\alpha e^{i2\theta}\rangle\nonumber\\
&+&\frac{\beta^{2}\gamma^{2}}{\sqrt{\gamma^{4}+2\beta^{4}}\sqrt{\gamma^{4}+\beta^{4}}}|H\rangle_{d1}|V\rangle_{b1}|H\rangle_{c1}|H\rangle_{e2}|\alpha
e^{i\theta}\rangle.\nonumber\\\label{combine5}
\end{eqnarray}
After the photons in the spatial modes $c1$ and $c3$ passing through
the QND2, if the homodyne measurement of the coherent state is
$\theta$,
  Eq. (\ref{combine5})
will also collapse to the maximally entangled W state with the same
form of Eq. (\ref{concentrated3}) or (\ref{concentrated4}), after
measuring the photon in the mode $e2$. The success probability
 $P_{2}^{2}$ of obtaining the states (\ref{concentrated3}) and (\ref{concentrated4})
includes two terms. Here the subscription "2" means the second
concentration round. The first term is the probability of the first
round to get the 0 (2$\pi$) phase shift. From Eq.( \ref{less20}), it
equals to
$\frac{\gamma^{4}+2\beta^{4}}{(\gamma^{2}+\beta^{2})(\gamma^{2}+2\beta^{2})}$.
The second term is the success probability to get the $\theta$ phase
shift
 in the second round. From Eq. (\ref{combine5}), it  equals to
$\frac{3\beta^{4}\gamma^{4}}{(\gamma^{4}+2\beta^{4})(\gamma^{4}+\beta^{4})}$.
Therefore, the whole success probability is
\begin{eqnarray}
P_{2}^{2}&=&\frac{\gamma^{4}+2\beta^{4}}{(\gamma^{2}+\beta^{2})(\gamma^{2}+2\beta^{2})}\frac{3\beta^{4}\gamma^{4}}{(\gamma^{4}+2\beta^{4})(\gamma^{4}+\beta^{4})}\nonumber\\
&=&\frac{3\beta^{4}\gamma^{4}}{(\gamma^{2}+2\beta^{2})(\gamma^{4}+\beta^{4})(\gamma^{2}+\beta^{2})}.
\end{eqnarray}

On the other hand, in the second concentration round, if the phase
shift is not $\theta$, but $0$ (2$\pi$), after measuring the photon
in the mode $e2$ by rotating $45^{\circ}$, the remaining state is
\begin{eqnarray}
&&\frac{\gamma^{4}}{\sqrt{\gamma^{4}+2\beta^{4}}\sqrt{\gamma^{4}+\beta^{4}}}|H\rangle_{d1}|H\rangle_{b1}|V\rangle_{c1}|H\rangle_{e2}\nonumber\\
&\pm&\frac{\beta^{4}}{\sqrt{\gamma^{4}+2\beta^{4}}\sqrt{\gamma^{4}+\beta^{4}}}|H\rangle_{d1}|V\rangle_{b1}|H\rangle_{c1}|V\rangle_{e2}\rangle\nonumber\\
&\pm&\frac{\beta^{4}}{\sqrt{\gamma^{4}+2\beta^{4}}\sqrt{\gamma^{4}+\beta^{4}}}|H\rangle_{d1}|V\rangle_{b1}|H\rangle_{c1}|V\rangle_{e2}\rangle.\nonumber\\\label{less3}
\end{eqnarray}
Compared with Eq. (\ref{less2}), state of Eq. (\ref{less3}) can also
be reconcentrated in a third round. In this way, this protocol can
be repeated for $N$ $(N\rightarrow\infty)$ times, in principle, if
each round can not get the $\theta$ phase shift.

 Through the above description, if they choose the $\theta=\pi$ phase shift,
the discarded items in Sec. II can be reused to get a higher success
probability. However, the natural cross-Kerr nonlinearity is
extremely small \cite{kok1,kok2}. It is hard to reach $\theta=\pi$.
Moreover, using longer interaction time will induce decoherence from
losses. It will make the output state become a mixed state.  A
practical alternative way is to use the coherent state rotation. We
take Eq. (\ref{combine3}) as an example. If they obtain the state
Eq. (\ref{combine3}),  Alice rotates the coherent state by $\theta$,
then Eq. (\ref{combine3}) becomes
\begin{eqnarray}
&\rightarrow&\frac{\alpha^{2}}{\sqrt{\alpha^{2}+\beta^{2}}}|V\rangle_{a1}|H\rangle_{a2}|H\rangle_{b1}|H\rangle_{c1}|\alpha e^{-i\theta}\rangle\nonumber\\
&+&\frac{\beta^{2}}{\sqrt{\alpha^{2}+\beta^{2}}}|H\rangle_{a1}|V\rangle_{a2}|V\rangle_{b1}|H\rangle_{c1}|\alpha e^{i\theta}\rangle\nonumber\\
&+&\frac{\alpha\gamma}{\sqrt{\alpha^{2}+\beta^{2}}}|H\rangle_{a1}|H\rangle_{a2}|H\rangle_{b1}|V\rangle_{c1}|\alpha \rangle\nonumber\\
&+&\frac{\beta\gamma}{\sqrt{\alpha^{2}+\beta^{2}}}|H\rangle_{a1}|V\rangle_{a2}|H\rangle_{b1}|V\rangle_{c1}|\alpha e^{i\theta}\rangle\nonumber\\
&+&\frac{\alpha\beta}{\sqrt{\alpha^{2}+\beta^{2}}}|V\rangle_{a1}|V\rangle_{a2}|H\rangle_{b1}|H\rangle_{c1}|\alpha \rangle\nonumber\\
&+&\frac{\alpha\beta}{\sqrt{\alpha^{2}+\beta^{2}}}|H\rangle_{a1}|H\rangle_{a2}|V\rangle_{b1}|H\rangle_{c1}|\alpha
\rangle.\label{combine6}
\end{eqnarray}
From Eq. (\ref{combine6}), if  there is no phase shift, the
remaining state is the same as Eq. (\ref{collpase1}), with the same
probability of $P^{1}$. Then following the same step described in
Sec. II, if D$_{1}$ fires, they will get
$|\Phi_{1}\rangle_{d1b1c1}$, and if D$_{2}$ fires, they will get
$|\Phi_{2}\rangle_{d1b1c1}$. Otherwise, they use the
$|X\rangle\langle X|$ homodyne detection \cite{QND1,QND2}  on the
coherent state \cite{he1,lin2}, which can make the $|\alpha
e^{i\theta}\rangle$ and $|\alpha e^{-i\theta}\rangle$ can not be
distinguished. Then the remaining state is the same as Eq.
(\ref{less0}). In this way, following the same step, it can be
reconcentrated to get a higher success probability. The way of
coherent state rotation can also be suitable for the second step  in
our the second protocol.

\section{Discussion and summary}
By far, we have fully explained our ECPs both with linear optics and
cross-Kerr nonlinearity. In order to explain ECPs clearly, for each
ECP we divide it into two steps. The first step is  operated by
Alice, and the second one is  operated by Charlie. Bob  needs only
to retain or discard his photons according to the measurement
results from Alice and Charlie by classical communication. In the
PBS protocol, the two steps should be performed simultaneously
because of the post-selection principle.  In the QND protocol, the
QND provides us a powerful tool to make quantum nondemolition
measurement which does not destroy the photons. This advantage makes
each step can be operated independently. Moreover, with QNDs, if
they choose the $\theta=\pi$, or use the coherent state rotation,
both  steps can be iterated to get a higher success probability. Now
let us calculate the success probability in each iteration rounds .

In the first step, we calculate the success probability in each
iterated  round as
\begin{eqnarray}
P_{1}^{1}&=&\frac{\alpha^{2}(\gamma^{2}+2\beta^{2})}{\alpha^{2}+\beta^{2}},\nonumber\\
P_{2}^{1}&=&\frac{\alpha^{4}(\beta^{2}\gamma^{2}+2\beta^{4})}{(\alpha^{4}+\beta^{4})(\alpha^{2}+\beta^{2})},\nonumber\\
P_{3}^{1}&=&\frac{\alpha^{8}(\beta^{6}\gamma^{2}+2\beta^{8})}{(\alpha^{8}+\beta^{8})(\alpha^{4}+\beta^{4})(\alpha^{2}+\beta^{2})},\nonumber\\
&\cdots&\nonumber\\
P_{N}^{1}&=&\frac{\alpha^{2^{N}}(\beta^{2^{N}-2}\gamma^{2}+2\beta^{2^{N}})}{(\alpha^{2^{N}}+\beta^{2^{N}})(\alpha^{2^{N-1}}+\beta^{2^{N-1}})\cdots(\alpha^{2}+\beta^{2})}.\nonumber\\
\end{eqnarray}
Here the superscription $"1"$ means the first step. The subscription
$"1","2","3",\cdots "N"$ is the iteration number.
\begin{figure}[!h]
\begin{center}
\includegraphics[width=7cm,angle=0]{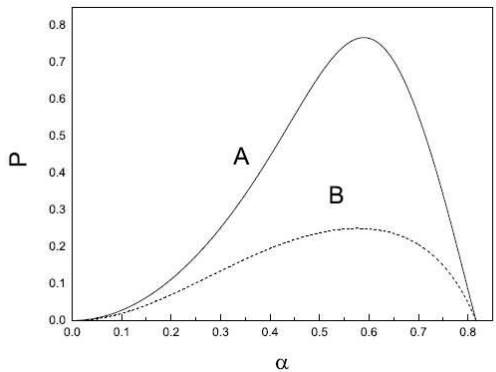}
\caption{The total success probability P of getting a maximally
entangled W state is altered with the initial coefficient $\alpha$.
Here, we choose $\beta=\frac{1}{\sqrt{3}}$,
$\alpha\in(0,\sqrt{\frac{2}{3}})$. Curve A is the QND protocol, and
Curve B is the PBS protocol. For numerical simulation, we choose
$N=M=3$ for approximation.}
\end{center}
\end{figure}

Following the same principle, in the second step, the  success
probability in each iterated round is
\begin{eqnarray}
P_{1}^{2}&=&\frac{3\beta^{2}\gamma^{2}}{(\gamma^{2}+\beta^{2})(\gamma^{2}+2\beta^{2})},\nonumber\\
P_{2}^{2}&=&\frac{3\beta^{4}\gamma^{4}}{(\gamma^{2}+2\beta^{2})(\gamma^{4}+\beta^{4})(\gamma^{2}+\beta^{2})},\nonumber\\
P_{3}^{2}&=&\frac{3\beta^{8}\gamma^{8}}{(\gamma^{2}+2\beta^{2})(\gamma^{8}+\beta^{8})(\gamma^{4}+\beta^{4})(\gamma^{2}+\beta^{2})},\nonumber\\
&\cdots&\nonumber\\
P_{M}^{2}
&=&\frac{3\beta^{2^{M}}\gamma^{2^{M}}}{(\gamma^{2^{M}}+\beta^{2^{M}})(\gamma^{2^{M-1}}+\beta^{2^{M-1}})\cdots(\gamma^{2}+\beta^{2})}\nonumber\\
&\cdot&\frac{1}{(\gamma^{2}+2\beta^{2})}.
\end{eqnarray}
Here the superscription $"2"$ means the second step. The
subscription $"1","2","3", \cdots "M"$ is also the iteration number.

With the QNDs, if the initial state is Eq. (\ref{W state}), the
whole concentration procedure can be described as follows: They
first perform the first step  first time. If it is a failure, then
they repeat  it
 again until it is successful. Then they go to the second step with the same iteration principle.
 Interestingly, in the first step, suppose  it is successful in the $K$th ($K=1,2\cdots
N)$ iteration, they always obtain  Eq. (\ref{concentrated1}), that
is
 the initial state of the second step.
Therefore, by repeating both steps, the total success probability is
\begin{eqnarray}
P_{total}&=&P_{1}^{1}(P_{1}^{2}+P_{2}^{2}+\cdots+P_{M}^{2})\nonumber\\
&+&P_{2}^{1}(P_{1}^{2}+P_{2}^{2}+\cdots+P_{M}^{2})\nonumber\\
&+&\cdots\nonumber\\
 &+&P_{N}^{1}(P_{1}^{2}+P_{2}^{2}+\cdots+P_{M}^{2})\nonumber\\
&=&\sum_{N=1}^{\infty}P^{1}_{N}\sum_{M=1}^{\infty}P^{2}_{M}.\label{total}
\end{eqnarray}
From Eq. (\ref{total}), the success probability of the PBS protocol
is essentially the first term of Eq. (\ref{total}), that is the case
of $N=M=1$. Interestingly, if
$\alpha=\beta=\gamma=\frac{1}{\sqrt{3}}$, the success probability of
the PBS protocol is $P=P_{1}^{1}
P_{1}^{2}=\frac{1}{2}\cdot\frac{1}{2}=\frac{1}{4}$. In this way for
the QND protocol,
\begin{eqnarray}
P_{total}&=&\sum_{N=1}^{\infty}P^{1}_{N}\sum_{M=1}^{\infty}P^{2}_{M}\nonumber\\
&=&(\frac{1}{2}+\frac{1}{4}+\frac{1}{8}+\cdots)\cdot(\frac{1}{2}+\frac{1}{4}+\frac{1}{8}+\cdots)\nonumber\\
&=& 1.
\end{eqnarray}
We calculate the total success probability of both the PBS and the
QND protocol, shown in Fig. 4.
 We choose
$\beta=\frac{1}{\sqrt{3}}$, and change
$\alpha\in(0,\sqrt{\frac{2}{3}})$. For the QND protocol, we choose
$N=M=3$ for a good numerical simulation. In Fig. 4, it is shown that
both
 success probability monotonic increase
 with $\alpha$, when $\alpha\in(0,\sqrt{\frac{1}{3}})$. They both have a
maximally value when $\alpha=\frac{1}{\sqrt{3}}$.

From Fig. 4, with QNDs,  one can get a higher success probability.
In this way, they should exploit the cross-Kerr nonlinearity to
generate $\pi$ phase shift on the coherent state, or use the
coherent state rotation. However, the largest natural cross-Kerr
nonlinearities are extremely weak ($\chi^{(3)}\approx
10^{-22}m^{2}V^{-2}$) \cite{kok1,kok2}.
 The kerr phase shift when operating in the optical single-photon
regime is about $\tau\approx10^{-18}$. With electromagnetically
induced transparent materials, it is much larger and can reach
$\tau\approx10^{-5}$.  On the other hand, using cross-Kerr
nonlinearity to implement the quantum information processing is
still a controversial topic \cite{he2,hofmann,
Banacloche,Shapiro1,Shapiro2,weak_meaurement}. In 2003, Hofmann
\emph{et al.} pointed out that a  $\pi$ phase shift can be reached
with a single two-level atom in a one-side cavity \cite{hofmann}. In
Ref. \cite{Shapiro1}, Shapiro argues that the single-photon Kerr
nonlinearities do not help quantum computation. Recently, He
\emph{et al.} discussed the cross-Kerr nonlinearity between
continuous-mode coherent state and single photons. They believed
that their work constitutes significant progress in making the
treatment of coherent state and single photon interactions more
realistic \cite{he2}.  Feizpour \emph{et al.} also showed that  it
is possible to amplify a cross-Kerr phase shift to an observable
value, which is much larger than the intrinsic magnitude of the
single-photon-level nonlinearity, with the help of weak measurement
\cite{weak_meaurement}. Giant cross-Kerr nonlinearities
 was also obtained with nearly
vanishing optical absorption, investigating the linear and nonlinear
propagation of probe and signal pules which is coupled in a double
quantum-well structure with a four-level, double $\Lambda$-type
configuration by Zhu and Huang \cite{oe}.

In summary, we have presented two ECPs for concentrating arbitrary W
states. We exploit both the linear optic element PBS and the
nonlinear optics cross-Kerr nonlinearity to achieve the whole task.
Compared with other concentration protocols, these protocols do not
require the collective measurement. Moreover, they do not need two
same copies of less entangled pairs to perform the protocol, which
make them be more economical.  In the PBS protocol, based on the
post-selection principle, one can obtain the maximally entangled W
state with certain probability. In the QND protocol, we adopt the
QNDs to substitute the PBSs and make the protocol become more
powerful. First, the parties can operate the protocol independently.
Second, it does not require the sophisticated single-photon
detectors. Third, by iterating this protocol, one can reach a higher
success probability. All these features maybe make these two
protocols more useful in  practical applications.

\section*{ACKNOWLEDGEMENTS}
This work was supported by the National Natural Science Foundation
of China under Grant No. 11104159, the Scientific Research
Foundation of Nanjing University of Posts and Telecommunications
under Grant No. NY211008, the University Natural Science Research
Foundation of JiangSu Province under Grant No. 11KJA510002, and the
open research fund of the Key Lab of Broadband Wireless
Communication and Sensor Network Technology (Nanjing University of
Posts and Telecommunications), Ministry of Education, China, and A
Project Funded by the Priority Academic Program Development of
Jiangsu Higher Education Institutions.

\end{document}